\documentclass[apl,twocolumn,aps,showpacs,10pt]{revtex4-2}

\usepackage{graphicx}  
\graphicspath{{./}}
\usepackage{dcolumn}  
\usepackage{amssymb, amsmath}
\usepackage{hyperref}
\usepackage{braket}

\begin{document}

\title{Long spin-1/2 noble gas coherence times in mm-sized anodically bonded batch-fabricated $^{3}$He-$^{129}$Xe-$^{87}$Rb cells}
\author{M.\ E.\ Limes}
\author{N.\ Dural}
\author{M.\ V.\ Romalis}
\affiliation{Department of Physics, Princeton University, Princeton, New Jersey, 08544, USA}
\author{E.\ L.\ Foley}
\author{T.\ W.\ Kornack}
\author{A. Nelson}
\author{L.\ R.\ Grisham}
\affiliation{Twinleaf LLC, Princeton, New Jersey, 08544, USA}
\date{\today}
\begin{abstract}

As the only stable spin-1/2 noble gas isotopes, $^{3}$He and $^{129}$Xe are promising systems for inertial rotation sensing and searches for exotic spin couplings. Spin-1/2 noble gases have intrinsic coherence times on the order of hours to days, which allows for incredibly low frequency error of free-precession measurements. However relaxation in miniature cells is dominated by interactions with the cell wall, which limits the performance of a chip-scale sensor that uses noble gases.  While $^{129}$Xe wall relaxation times have previously been limited to 10's of seconds in mm-sized cells, we demonstrate the first anodically bonded batch-fabricated  cells with dual $^{3}$He-$^{129}$Xe isotopes and $^{87}$Rb in a 6~mm$^3$ volume with $^{3}$He and $^{129}$Xe $T_2$ coherence times of, respectively, 4 h and 300 s. We use these microfabricated cells in a dual noble-gas comagnetometer and discuss its limits.
 
\end{abstract}

\maketitle
%

The research and commercial development of minature high-performance atomic sensors  \cite{Liew_2004,Knappe_2006,Kitching_2018, Limes_2020,McGilligan_2020} has naturally driven interest in chip-scale systems for hyperpolarization of noble gas nuclear spins \cite{Ledbetter_2008,Jimenez_2014}. Compact noble-gas NMR systems are used in a variety of studies, including searches of short-range spin-dependent forces \cite{Bulatowicz_2013}, inertial rotation sensing \cite{Donley_2010,Larsen_2012,Walker_2016,Karlen2018,Thrasher_2019}, magnetometry \cite{LarsenMag}, and microfluidic detection \cite{Pines_2014,Kennedy_2017}. These applications use vapor cells that contain the readily available $^{129}$Xe or $^{131}$Xe that are hyperpolarized by spin exchange with optically pumped alkali metals. Noble-gas nuclei are notable for having extraordinarily long $T_2$ coherence times while having high densities, making the quantum projection frequency noise $\delta \omega = 1/\sqrt{T_2 N t}$ well below pHz levels \cite{Gentile_2017}, with $N$ the number of nuclei sampled and measurement time $t$. The potential of noble gases for rotation measurements has long been considered \cite{Kastler_1973,Lam_1983,Mehring_1992}, but more than one simultaneous measurement is needed to discern rotations of the apparatus from changes in the local magnetic field. Thus the use of a comagnetometer that samples the same magnetic field in the same volume is advantageous for rotation measurements done in a  compact and cost-effective package. In principle the spin-1/2 isotopes $^{129}$Xe or $^{3}$He \cite{Chupp_1987, Stoner_1996, Bear_2000} are the best candidates for systematic-free rotation measurements, as higher-spin isotopes like $^{131}$Xe or $^{21}$Ne are susceptible to additional frequency shifts and relaxation mechanisms that can be deleterious \cite{Chupp_1985,Wu_1987,Chupp_1988,Butscher_1994, Donley_2009}. Detection of nuclear spins at a quantum shot noise level is difficult to achieve. A popular and sensitive detection method that takes advantage of the wavefunction overlap of the optically addressable electronic state of an alkali metal with a noble gas nuclei  \cite{Walker_1997, Appelt_1998}. While backaction and relaxation from the alkali on the noble gas must be mitigated \cite{Limes_2017, Korver_2015}, this method of detection is convenient because Rb alkali vapor is already present in these systems to hyperpolarize $^{3}$He and $^{129}$Xe through spin-exchange optical pumping. 

Spin-1/2 noble gases intrinsically decohere very slowly through intraspecies collisional processes that form transient or persistent dimers which allow for dipole-dipole and spin-rotation relaxation \cite{Newbury_1993, Chann_2002, Anger_2008}. Various extrinsic factors that cause nuclear relaxation include diffusion through magnetic field gradients, spin exchange and spin destruction of the noble gases with warm alkali vapors, and relaxation due to the cell wall.  Given a particular coating or cell surface, the characteristic wall relaxation time $T_{\text{wall}}$ scales linearly with surface-to-volume ratio and follows an Arrhenius  temperature dependence. In small cells with a large surface-to-volume ratio, the noble-gas coherence times are often limited by interactions with the cell walls, with past studies showing $^{129}$Xe relaxation of 10's of seconds \cite{Walker_2016}. We demonstrate batch fabrication of stemless anodically bonded cells that have a well-defined geometry and contain $^{3}$He  and $^{129}$Xe with nuclear-spin coherence times of 4 hours and 300 s, respectively, a significant improvement in the $^{129}$Xe coherence time for micro-fabricated cells and the first successful detection of $^3$He signals in such cells. We then use these cells as a free-precession dual noble-gas comagnetometer in a laboratory-sized prototype for a chip-scale gyroscope, achieving a stability of 0.2 deg/h at 48 h. 
\begin{figure}
\includegraphics[width=3in]{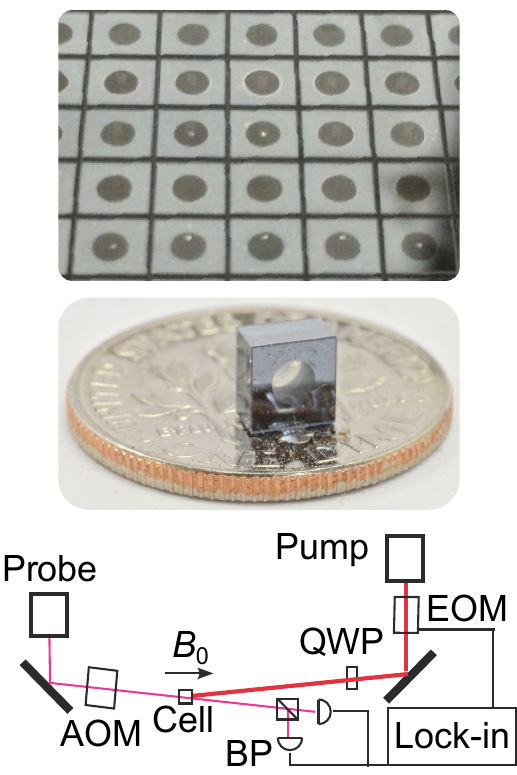}
\caption{ Drilled silicon wafers can be filled with a variety of noble gases and enriched alkali isotopes then anodically bonded and diced to make many uniform batch fabricated vapor cells. We fill cells with $^3$He-$^{129}$Xe-$^{87}$Rb-N$_2$ for use in a dual noble-gas comagnetometer detected by an in situ alkali magnetometer and use long noble-gas relaxation times to study a low-drift chip-scale gyroscope. For detection of the noble gases, we use co-linear probe and pump beams, with pump helicity flipped by an electro-optical modulator (EOM) and probe shuttered with an acousto-optical modulator (AOM), along with a fast $^{87}$Rb $\pi$ pulse train applied with small Helmholtz coils. }
\label{fig:1}
\end{figure}

For this study we make stemless cells shown in Fig.~\ref{fig:1} by using a silicon wafer of thickness 2 mm, drilled with an array of 2 mm holes, machined with two 0.2 mm thick SD-2 aluminosilcate glass plates on each side \cite{Liew_2004,Kitchingglass}.  The wafer and glass are baked under high vacuum, which is crucial to obtain long $^{129}$Xe lifetimes.  We first bond glass to one side of the wafer, and then introduce $99.9\%$ isotopically enriched $^{87}$Rb metal into the cells. Then the other side of the wafer is bonded to enclose the cell under a 1400:6.5:80 torr $^{3}$He:$^{129}$Xe:N$_2$ gas mixture in the bonding chamber. The $N_2$ is included for quenching during optical pumping of $^{87}$Rb. Heat applied during bonding can cause buffer gas pressure loss in cells of up to 30$\%$; we check the $^{3}$He and $^{87}$Rb densities by measuring the Rb absorption spectrum of the probe laser. A cryogenic storage system reclaims $^{3}$He from the bonding chamber, enabling a pathway for fabricating many low-cost devices. Anodically bonded glass on machined silicon also allows for excellent control of the cell geometry. Typical glass blown or optically contacted cells have a glass stem for cell filling and sealing \cite{Eklund_2008,Larsen_2012}, which creates an inherent asymmetry in the cell shape. Some are able to lessen cell stem effects by plugging the stem with alkali metal \cite{Lee_2018} or a small piece of movable glass \cite{Balabas_2010}. Others have made stemless cells by allowing  $^3$He to diffuse through quartz walls at a high temperature \cite{Heil_2016}.  The cell shape is an important consideration for precision measurements and chip-scale sensors that use noble-gas precession frequency measurements, as it affects the long-range dipolar magnetic interactions between spin-polarized nuclei \cite{Nacher_1995}. Such interactions cause significant frequency shifts in precision comagnetometer measurements  \cite{Allmendinger_2014}, where long-term drifts can be correlated to placement and magnitude of noble-gas polarization \cite{Romalis_2014, Limes_2019, Vaara_2019, Terrano_2019,Sachdeva_2019}. To study this dipolar effect, we recently used our cell fabrication process to make a slight wedge in the Si wafer to vary the height-to-diameter aspect ratios across many cells. We demonstrated the existance of an optimal aspect ratio that nulls the dipolar effect, as well as the existence of nuclear $J$-coupling between different noble gases $^{3}$He-$^{129}$Xe \cite{Limes_2019}. This $J$-coupling in principle also provides a mechanism for relaxation and polarization transfer between the different noble gas species, although this has not yet been directly observed. 

We include $^{87}$Rb in the cells in order to polarize nuclear spins by spin-exchange with $^{87}$Rb that is polarized by optical pumping using a circularly polarized laser beam parallel to a longitudinal bias magnetic field $B_0 = 0.5$ {\textmu}T. 
We also detect noble-gas spin precession with a $^{87}$Rb magnetometer \cite{Sheng_2014, Limes_2017, Zhivun_2019}, which gives a high signal-to-noise ratio due to wavefunction overlap of the $^{87}$Rb with the noble gases \cite{Schaefer_1989,Romalis_1998,Ma_2011, Nahlawi_2019}. 
Since the cells have only a single optical axis, counter-propagating pump and probe beams are used with fast $^{87}$Rb magnetic field $\pi$ pulses to make a pulse-train magnetometer similar to that described in Ref.~\cite{Limes_2017, Limes_2019}. For $^{87}$Rb magnetometer operation, the polarization of an 795 nm pump laser is alternated between $\sigma^+$ and $\sigma^-$ light at 13 kHz by an EOM. Synchronous 2 $\mu$s long magnetic field pulses along the $y$-axis (or any axis in the transverse plane) quickly Rabi flop the $^{87}$Rb polarization back and forth along $B_0$. 
If the pulses are sufficiently fast and short, they can suppress alkali-alkali spin-exchange relaxation as well as precession of $^{87}$Rb spins around $B_0$ \cite{Limes_2017}. The $^{87}$Rb polarization along $B_0$ is detected with a probe beam that is detuned off the $D1$ resonance, passing through the cell to a balanced polarimeter. The probe laser is shuttered by an AOM to open only during $\pi$ pulses to mitigate probe broadening. A DC or low frequency $B_y$ field parallel to the direction of the fast pulses delays or advances the $^{87}$Rb signal zero-crossing during the $\pi$ Rabi flop.   This signal is measured by a lock-in amplifier referenced to half the EOM frequency with the lock-in output proportional to $B_y$.  The sensitivity of the $^{87}$Rb magnetometer is 300 fT/$\sqrt{\text{Hz}}$. We use sub-mW power from Photodigm DBR lasers through the active volume, similar to the powers from commercial VCSEL laser packages. 

\begin{figure}
\includegraphics[width=3in]{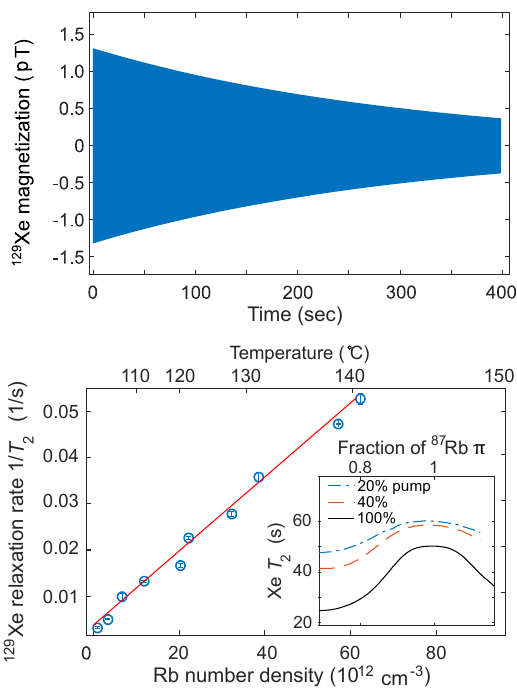}
\caption{Top: Spin precession signal from $^{129}$Xe at 73.3$^{\circ}$C show relaxation times of $T_2 = 308$ s. Bottom: The dependence of $^{129}$Xe $T_2$ on the Rb number density. The inset shows that at $120^{\circ}$C  $^{87}$Rb pump light intensity and deviations of the alkali magnetometer $\pi$ pulse amplitude from optimal conditions shorten $^{129}$Xe $T_2$ due to Xe diffusion in a $^{87}$Rb polarization gradient.}
\label{fig:2}
\end{figure}

Although self-relaxation for the noble gases in our conditions yields coherence times of hundreds of hours for both $^{129}$Xe and $^{3}$He, our system is dominated by wall relaxation and Rb-noble gas interactions. 
To study $^{129}$Xe relaxation mechanisms, we first spin-exchange optically pump the $^{129}$Xe along the bias field for 30-50 s, and have negligible $^{3}$He polarization. 
A tipping pulse then places  $^{129}$Xe transverse to $B_0$ and its precession is detected with the pulse-train $^{87}$Rb magnetometer. A representative signal from the lock-in amplifier is shown in Fig.~\ref{fig:2}a. We fit a given data set to the function $A\exp(-t/T_2)\sin(\omega_{\text{Xe}} t)$ to extract $T_2$, and show  $^{129}$Xe $1/T_2$  as a function of cell temperature  and Rb density  in Fig.~\ref{fig:2}b.  In this cell we find the Rb absorption  FWHM of 23.4~$\pm 1$~GHz corresponding to  $1000$~torr $ ^{3}$He \cite{Romalis_1997}. From the slope in Fig.~\ref{fig:2}b  we find a Rb-Xe spin-exchange rate of $(7.8 \pm 0.7) \!\times\! 10^{-16}$ cm$^3$/s. This value is in agreement with a calculation of the spin-exchange rate $7.2 \times 10^{-16}$ cm$^3$/s based on previously measured cross-sections \cite{Nelson_2001, Schrank_2009}. 
%
The relaxation rate does not change significantly from the Rb-Xe spin-exchange rate across our temperature range, so we neglect any Arrhenius temperature dependence in our analysis. The inset of Fig.~\ref{fig:2}b shows the Rb magnetometer can shorten $ ^{129}$Xe $T_2$ by causing a Rb polarization gradient that $^{129}$Xe diffuses through.  The additional relaxation depends on the $^{87}$Rb pump laser  power and the accuracy of  $^{87}$Rb $\pi$ pulses.  Increasing the $^{87}$Rb $\pi$ pulse repetition rate  decreases this relaxation   because the Rb gradient is reversed more rapidly. For accurate $T_2$ measurements we use proper $\pi$ pulses and low pump power. The $^{3}$He coherence exceeds $^{129}$Xe $T_2$ by at least an order of magnitude, thus sufficient for our comagnetometer operation. 

\begin{figure}
\includegraphics[width=3in]{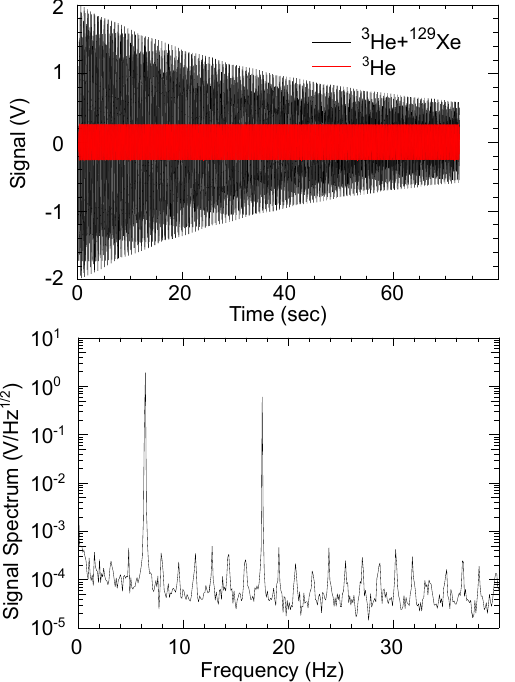}
\caption{Top: Time-domain $^{3}$He-$^{129}$Xe free-precession comagnetometer signals detected with an in situ $^{87}$Rb magnetometer at an operating temperature of 120$^\circ$C. Bottom: Spectral density of comagnetometer in mm-scale cells. }
\label{fig:3}
\end{figure} 

\begin{figure}
\includegraphics[width=3in]{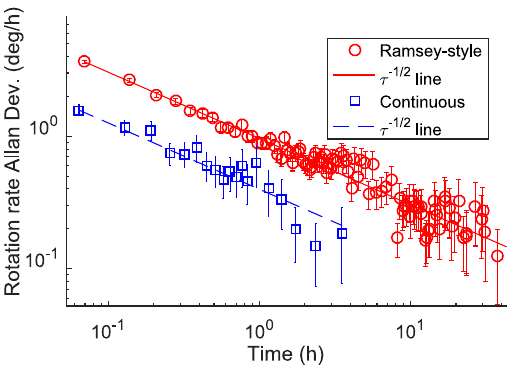}
\caption{Our system stability is shown with Allan deviations for $^{3}$He-$^{129}$Xe comagnetometer detected with an in situ $^{87}$Rb magnetometer. We compare a continuous-mode pulsed magnetometer operation with a Ramsey-style sequence that has an Rb-noble gas decoupling and dark period separating two detection periods. Both data sets are taken in a batch-fabricated anodically bonded cell at 120$^{\circ}$C, and have a practical limit of achievable steady-state $^{3}$He polarization.}
\label{fig:4}
\end{figure} 

To study the feasibility of these long noble gas lifetime cells in a compact gyroscope we operate the noble gases in a free-precession comagnetometer, similar to measurements in Ref.~\cite{Limes_2017} that were with larger glass-blown cells. For these measurements we hyperpolarize $^{3}$He and $^{129}$Xe along a bias field $B_0 = 0.5$ {\textmu}T that is co-linear to the pump and probe beams. We then apply a tipping pulse that places $^{3}$He and $^{129}$Xe spins in a plane transverse to $B_0$. 
The $^{87}$Rb pulse-train magnetometer detects the precession of both noble gases about $B_0$ as shown in Fig.~\ref{fig:3}.  To find the rotation rate for a given shot with both measurement styles, we first find the effective free-precession frequencies of the noble gases $\omega_{\text{He}} = \gamma_{\text{He}}B_0 + \Omega$ and $\omega_{\text{Xe}} = \gamma_{\text{Xe}}B_0 + \Omega$. We extract the rotation rate $\Omega$ by taking the ratio $f_r = \omega_{\text{He}}/\omega_{\text{Xe}}$ and finding $\Omega = \omega_{\text{Xe}}(\gamma_r- f_r)/(\gamma_r - 1)$ with $\gamma_r = \gamma_{\text{He}}/\gamma_{\text{Xe}} = 2.7540813(3)$ \cite{Makulski_2015}. 
After a measurement the noble gas spins are placed back along the bias field by dumping feedback from the lock-in signal \cite{Alem_2013}. Repeating this sequence, we run the comagnetometer for an extended period of time to determine its sensitivity and long-term stability. In Fig.~\ref{fig:4} we compare the Allan deviations for a Ramsey-style sequence with active Rb depolarization and for measurements with the Rb $\pi$-pulse train magnetometer continuously running. 
For the Ramsey-style measurement each shot takes about 115 s, with a 25 s inital and 45 s final detection period. We fit the detection period signals to decaying sine waves to determine the phases $\phi_1$ and $\phi_2$ with which the spins enter and leave the dark detection period $T_d$ = 45 sec. Assuming there are no significant changes in the precession frequency, we find the in-the-dark free-precession frequencies of the noble gases by taking $\omega = \Delta \phi / \Delta t = (2n\pi + \phi_2 - \phi_1)/T_d$, where $n$ is an integer determined from an estimate of the precession frequency. During the dark period a two-axis decoupling pulse train is used to mitigate noble-gas frequency shifts due to $^{87}$Rb back polarization and nulls Bloch-Siegert shifts introduced by the pulse train \cite{Limes_2017}. 
For continuous measurements each shot lasts about 50 s, and we fit two decaying sine waves to find $\omega_{\text{Xe}}$ and $\omega_{\text{He}}$. In both measurement styles we find frequency uncertainty is limited by the $^{3}$He frequency error due to low steady-state $^{3}$He magnetization at a temperature that allows for long $^{129}$Xe $T_2$, as higher temperatures cause greater Rb-Xe relaxation. The Ramsey-style measurement integrates for more than a day, longer than the continous method by about an order of magnitude, with both methods reaching around 0.2 deg/h precision before drifting. 
The uncertainties achieved here are about an order of magnitude worse than achieved in the same system for glass-blown cells with 100 times larger volume used in Ref.~\cite{Limes_2017}. 
Using microfabricated cells with higher $^3$He pressure would improve the performance, as in the current cells the small steady-state magnetization becomes insufficient with practical shot-to-shot pumping periods. 
To determine expected gyroscope performance we can use Cram\'er-Rao lower bounds (CRLB) of a decaying sine wave \cite{Chupp_1994,Yao_1995,Gemmel_2010} to estimate a lower bound for the rotation error 
\begin{equation}
\delta \Omega = \frac{1}{\gamma_r - 1}\sqrt{ \gamma_r^2 \delta\omega_{\text{Xe}}^2+ \delta\omega_{\text{He}}^2}.
\label{eq:rotEst}
\end{equation}
For a decaying sine wave, 
\begin{equation}
\begin{split}
\frac{\delta \omega}{2 \pi} &= \frac{\sqrt{12C}}{2\pi T_m^{3/2} \text{SNR}} ~[\text{Hz}] , \\
\delta \phi &= \frac{2\sqrt{ D }}{ T_m^{1/2} \text{SNR}} ~[\text{rad}], 
\end{split}
\end{equation}
where $T_m$ is the measurement time and SNR $=A/\rho$, where $A$ is the initial sine wave amplitude and $\rho$ is the spectral density of white Gaussian noise. $C$ and $D$ are factors depending on the measurement and relaxation time $T_m$ and $T_2$ that both approach unity as $T_2 \rightarrow \infty$. 
For Ramsey-style detection in the limit of long $T_2$, the optimal detection intervals are $1/6$ of the total measurement time and the overall statistical sensitivity is three times worse than for continuous detection of equal time. 
For finite relaxation one can show that 
\begin{equation}
\begin{split}
C &= \frac{{T_m}^3 \left(e^{\frac{2 {T_m}}{{T_2}}}-1\right)}{3
   {T_2}^3 \cosh \left(\frac{2 {T_m}}{{T_2}}\right)-3
   \left({T_2}^3+2 {T_2} {T_m}^2\right)}, \\
   D_1&=\frac{e^{\frac{2 T_m}{T_2}} T_m \left(-2 T_2 T_m+2
   T_m^2+T_2^2\right)-T_2^2 T_m}{2 T_2^3 \cosh \left(\frac{2
   T_m}{T_2}\right)-2 \left(2 T_2 T_m^2+T_2^3\right)},\\
   D_2&=\frac{T_2^2 T_m \left(e^{\frac{2 T_m}{T_2}}-1\right) -2 T_m^3-2 T_2 T_m^2}{2 T_2^3 \cosh \left(\frac{2
   T_m}{T_2}\right)-2 \left(2 T_2 T_m^2+T_2^3\right)},
\end{split}
\end{equation}
where $D_1$ is used to for estimate of the trailing phase of the first detection period and $D_2$ for the leading phase of the second period. 
Using the signal height, noise, and relaxation times from Fig.~\ref{fig:3}, and considering detection, in-the-dark, and shot-to-shot times used in Fig.~\ref{fig:4} we find a Cram\'er-Rao frequency uncertainty lower bound that corresponds to an angular random walk of 0.094 deg/$\sqrt{\text{h}}$ for a Ramsey-style measurement and 0.056 deg/$\sqrt{\text{h}}$ for continuous operation, about an order of magnitude better than we realized experimentally.


In conclusion, we demonstrate fabrication of miniature stemless anodically bonded vapor cells containing $^{3}$He, $^{129}$Xe, $^{87}$Rb, and N$_2$ with long $^{129}$Xe  and $ ^{3}$He coherence times. We use these cells in a dual noble-gas comagnetometer detected by an in situ $^{87}$Rb pulse-train magnetometer, and study the long-term stability of the system. Though far from fundamental limits, this system has great potential for high-precision, accurate rotation measurements in a small, low-cost package. 

This  work  was  funded  by  DARPA  (Defense  Advanced Research Project Agency) under contract FA8650-13-1-7326 and NSF Grant No. 1404325.

\end{document}